\documentclass[11pt]{article}
\usepackage{hyperref}
\usepackage[final]{graphicx}
\usepackage{float}
\usepackage{amsmath}

\pdfoutput=1
\begin{document}
\title{Hydrodynamics in the \\ wake of a pitching foil}
\author{R. Fernandez-Prats \& F. J. Huera-Huarte \\
\\\vspace{6pt} Department of Mechanical Engineering, \\ Universitat Rovira i Virgili, 43007 Tarragona, Spain}
\maketitle
\begin{abstract}
fluid dynamics videos. The effect of flexibility on the hydrodynamic loads and on the flow structures generated on a rectangular foil when oscillating in pitch has been studied. Hydrodynamic loads were measured with a 6-axes balance, and the flow structures were investigated by using a Digital Particle Image Velocimetry ($DPIV$). It is known from nature's fin based propulsion mechanisms, that appendage stiffness plays an important role in their propulsive efficiency. We have studied four different stiffnesses, ranging from completely rigid to highly flexible. Optimal efficiency has been observed for an intermediate case. In this case, a moderately stronger trailing-edge vortex system takes place. A very high level of flexibility of the foil results in a reduction of efficiency.
\end{abstract}

\section{Explanation of the video}
 \subsection{Experimental system}

\begin{figure}[ht]
\centering
\includegraphics[bb=0 0 928 415, scale=0.45]{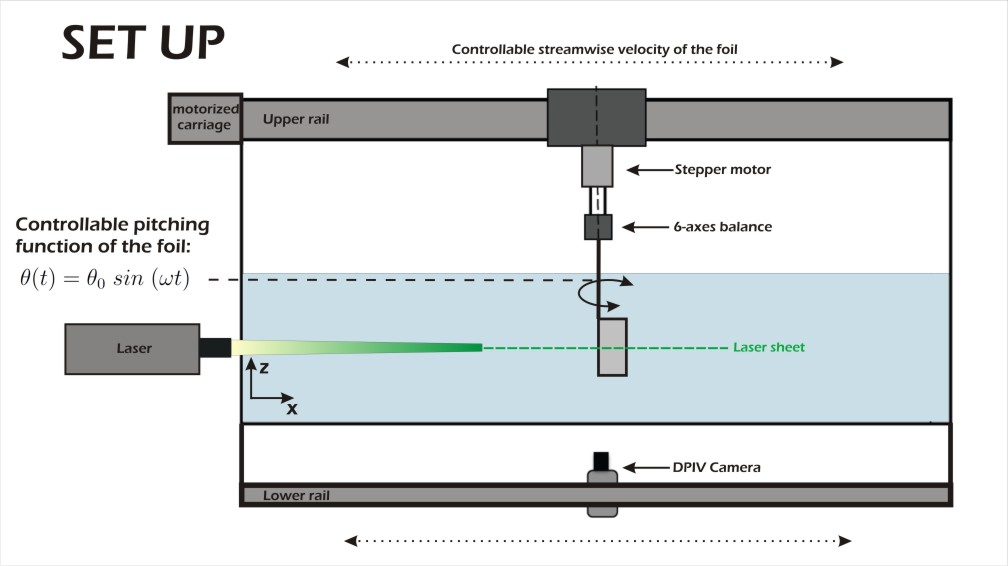} \caption{Experimental setup and devices.} \label{fig:Imageset}
\end{figure}

Experiments were conducted in the towing carriage of the Laboratory for Fluid-Structure Itneraction (LIFE), at the the Department of Mechanical Engineering of the Universitat Rovira i Virgili. The water tank is 2 m long and it has a cross section of $0.6 x 0.6 m^2$. Towing speeds in the carriage can be accurately controlled, see figure \ref{fig:Imageset} for details.

\subsection{Apparatus}
The steel foil has a chord of $c$ = 11 cm and a vertical span of $s$ = 20 cm. It is atached to a shaft driven a stepper motor which can be accurately controlled at the same time as the towing speed. The mean vertical position of the plate is 15 cm below the water surface and 15 cm above the floor of the tank. A balance is used to measure lift and drag forces, as well as pitch moment \ref{fig:Imageset}. The pitch has equation:

\begin{equation}
\centering
\begin{aligned}
\theta(t)=\theta_{0}\; sin\;(\omega t),
\end{aligned}
\end{equation}

where $\omega$ is the frequency of the oscillation and $\theta_{0}$
is the amplitude of the pitch motion.

\subsection{Techniques applied}

Planar $DPIV$ data was obtained after illuminating a region around the foil using a planar laser sheet generated by a Diode-pumped Solid State device. The light scattered by the tracer was captured using a digital camera based on a 4 Mpixel CMOS sensor. The flow  was seeded using  \(  10 \ \mu \)m diameter Silver Coated Hollow Glass Spheres (S-HGS).

\vspace{0.1cm}

$Fx$ thrust, $Fy$ lift and $Mz$ pitch torque were sampled at a frequency of 5kHz. $DPIV$ and load measurements were taken synchronously, so they can be related to each other easily.
\begin{figure}[ht]
\centering
\includegraphics[bb=0 0 828 435, scale=0.35]{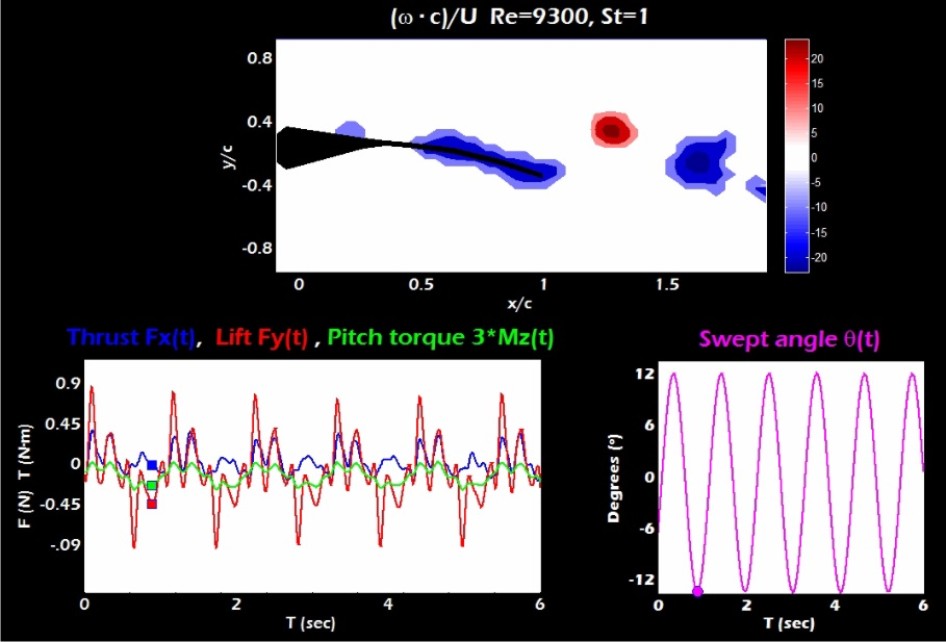} \caption{$DPIV$ (figure at the top) and load analysis (Thrust, lift and pitch torque, figure at the left) Swept angle of the foil at the right figure.} \label{fig:ImagePIVF}
\end{figure}

\subsection{Parameters, Equations and Coefficients}

Our set up installed at the towing tank, allowed the control of the flapping amplitude (A), frequency (f), as well as the free-stream or towing velocity (U). Four different flexibilities have been studied, rigid foil, semi-rigid foil, semi-flexible foil and flexible foil. A total of 720 experimental runs have been done with the techniques described above as you can see at the figure \ref{fig:ImagePara}. The following definitions apply:
\begin{figure}[ht]
\centering
\begin{minipage}[b]{0.48\linewidth}
\includegraphics[bb=0 0 1128 375, scale=0.25]{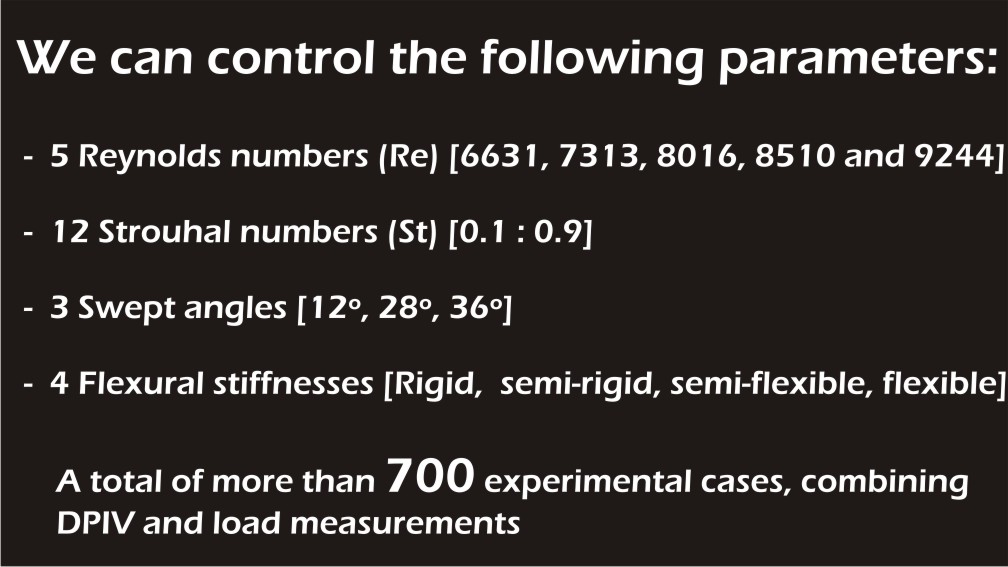} \caption{Parameters.} \label{fig:ImagePara}
\end{minipage}
\quad
\begin{minipage}[b]{0.48\linewidth}
\includegraphics[bb=0 0 1128 375, scale=0.25]{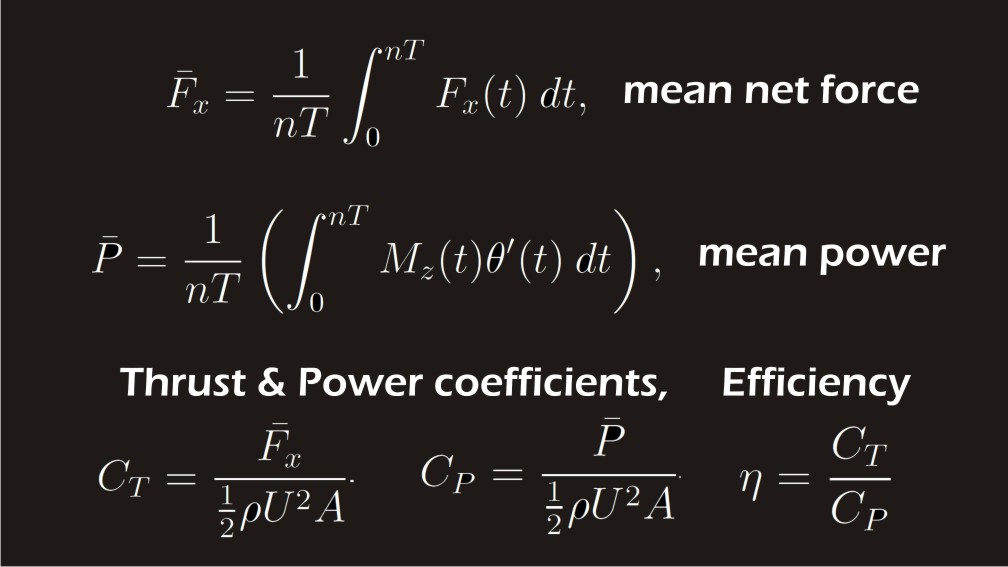} \caption{Equations and coefficients.} \label{fig:Imageecu}
\end{minipage}
\end{figure}

 \subsection{Result}

As already mentioned, the flexibility plays an important role in propulsion. We have found out that the optimal efficiency is produced for an intermediate case of flexibility. Efficiency is higher for two main reasons:

\begin{enumerate}

\item Thrust coefficient increases with increasing flexibility until a certain optimum point.

\item Power coefficient decreases with increasing flexibility until a certain optimum point.

 \end{enumerate}


\begin{thebibliography}{4}



\bibitem{Ander} {\sc{Anderson, J. M., Streitlien, K., Barrett, D. S. \& Triantafyllou, M. S. 1998.}} Oscillating foils of high propulsive effciency. J. Fluid Mech. 360, 41-72.

\bibitem{Oscillat} {\sc{    Triantafyllou, M. S., Triantafyllou, G. S. \& Yue, D. K. P. 2000.}} Hydrodynamics of Fishlike Swimming. Annu. Rev. Fluid Mech. 32, 33-53.


\bibitem{Huerawake1} {\sc{ F.J. Huera-Huarte, P.W. Bearman.}} Wake structures and
vortex-induced vibrations of a long flexible cylinder - Part 1:
Dynamic response. J Fluid Struct, 25, 969-990, 2009.
[doi:10.1016/j.jfluidstructs.2009.03.007]


\bibitem{raffel} {\sc{M. Raffel C. Willert S. Wereley J. Kompenhans}}
Particle Image Velocimetry. Second Edition, Springer.


\end{thebibliography}
\end{document}